\documentclass[12pt,preprint]{aastex}

\usepackage{graphicx}   

\shorttitle{GRB Energetics in the Swift Era}
\shortauthors{Kocevski et al.}

\begin{document}

\title{GRB Energetics in the Swift Era}

\author{Daniel Kocevski \altaffilmark{1}, Nathaniel Butler \altaffilmark{1} }

\altaffiltext{1}{Astronomy Department, University of California, 601 Campbell Hall, Berkeley, CA 94720 }
\email{kocevski@berkeley.edu, nat@astro.berkeley.edu}


\begin{abstract}

We examine the rest frame energetics of 76 gamma-ray bursts (GRBs) with known redshift that were detected by the Swift spacecraft and monitored by the satellite's X-ray Telescope (XRT).  Using the bolometric fluence values estimated in \citet{Butler07b} and the last XRT observation for each event, we set a lower limit the their collimation corrected energy $E_{\gamma}$ and find that a 68$\%$ of our sample are at high enough redshift and/or low enough fluence to accommodate a jet break occurring beyond the last XRT observation and still be consistent with the pre-Swift $E_{\gamma}$ distribution for long GRBs.  We find that relatively few of the X-ray light curves for the remaining events show evidence for late-time decay slopes that are consistent with that expected from post jet break emission.  The breaks in the X-ray light curves that do exist tend to be shallower and occur earlier than the breaks previously observed in optical light curves, yielding a $E_{\gamma}$ distribution that is far lower than the pre-Swift distribution.  If these early X-ray breaks are not due to jet effects, then a small but significant fraction of our sample have lower limits to their collimation corrected energy that place them well above the pre-Swift $E_{\gamma}$ distribution.  Either scenario would necessitate a much wider post-Swift $E_{\gamma}$ distribution for long cosmological GRBs compared to the narrow standard energy deduced from pre-Swift observations.  We note that almost all of the pre-Swift $E_{\gamma}$ estimates come from jet breaks detected in the optical whereas our sample is limited entirely to X-ray wavelengths, furthering the suggestion that the assumed achromaticity of jet breaks may not extend to high energies.

\end{abstract}


 \maketitle


\section{Introduction} \label{sec:Introduction}
 
As of June 2007, the Swift spacecraft's Burst Alert Telescope (BAT) \citep{Gehrels04,Barthelmy05a} had detected over 200 gamma-ray bursts (GRBs) and has followed $>80\%$ with the satellite's X-ray Telescope (XRT) \citep{Burrows05a}.  The data that has accumulated as a result of the huge success of the XRT has shown that the X-ray light curves of GRB afterglows are far more complex than previous observations \citep[e.g.,][]{Frontera00, Gendre06} had indicated.  Large drops in the X-ray emission immediately following a GRB \citep{Barthelmy05b} are superseded by a shallow decay \citep{Granot06}, ultimately giving way to the late-time afterglow light curve observed by pre-Swift X-ray instruments.  In many cases these light curve phases are punctuated by flaring activity occurring hundreds to thousands of seconds after the initial energy release \citep{Burrows05b}.  Yet in other cases only a single light curve phase manifests, yielding an uninterrupted power law decline that extends directly from the prompt gamma-ray emission into the X-ray regime lasting several days to weeks after the event \citep{Schady06,Sato07,Mundell07,Holland07}.  

Relatively few of these XRT monitored afterglow light curves have shown properties consistent with the  late-time steepening that had been observed to occur in the optical light curves of pre-Swift GRBs \citep{Harrison99,Stanek99}.  This sharp drop in the flux of some pre-Swift afterglows has been interpreted as a sign of the deceleration and/or lateral expansion of a highly collimated relativistic outflow \citep{Rhoads97}.  The existence of such a jet structure in the GRB outflow has become an integral part of the theoretical description of these events \citep{Meszaros02} and indeed a necessary component to explain the enormous amount of radiated energy inferred if the prompt emission is assumed to be isotropic \citep{Waxman98,Fruchter99}. 

Several authors have examined the presence, or lack thereof, of jet breaks in the X-ray afterglow light curves collected by the XRT. \citet{Burrows07} examined the X-ray light curves of $\sim$150 GRB afterglows and concluded that the ``canonical" jet model behavior, consisting of an achromatic light curve break between $t_{\rm jet} \sim$ 1 to 4 days and a post-break power law decay index $F_{\nu} \propto t^{-\alpha}$ steeper than $\alpha = 2.0$ \citep{Rhoads97,Sari99}, are extremely rare.  They find that many of these X-ray light curves do exhibit breaks, but that they occur at about $10^{4}$ seconds, far earlier than the jet breaks observed at optical wavelengths of pre-Swift GRBs.  These early breaks are typically followed by power law decays which are shallower than the minimum decay index predicted by simple jet models.  In all, the preliminary analysis done by \citet{Burrows07} revealed only 6 events with light curve breaks that were consistent with theoretical predictions of jet break behavior.  Subsequently, \citet{Panaitescu07a} performed a similar analysis of 236 GRB afterglows and found 30 events which were consistent with the behavior expected from standard jet models.  He also reported an additional 27 events with potential jet breaks, for which the spectral and temporal properties were not entirely consistent with model predictions, and another 38 events which exhibit no temporal breaks in their X-ray light curves.  In all, Panaitescu concludes that some 60$\%$ of well-monitored X-ray afterglows exhibit some evidence for a potential late-time jet break.

Recently, \citet{Butler07b} reported on the first comprehensive catalog of bolometric energy fluences of GRBs detected by the BAT instrument.   One implication of that analysis is that a a significant fraction of the Swift events with known redshift $z$ are under-energetic relative to pre-Swift events.  This is likely due to a factor 3--10 greater senstivity and lower resulting detection threshold of the BAT, relative to previous instruments. \citep[e.g.,][]{Barthelmy05a}. One consequence of this higher sensitivity may be a capacity to detect a greater fraction of more distant, high-$z$, events.  Both of these effects, a lower $E_{\rm iso}$ and a higher $z$, have the effect of increasing the predicted jet break time $t _{\rm jet}$ given a fixed collimation angle, or more importantly, a standard collimation corrected energy $E_{\gamma}$.

In this paper, we examine the source frame energetics of 76 GRBs detected by the Swift spacecraft with known redshift.  Using the bolometric fluence values estimated in \citet{Butler07b} and the last XRT observation, we calculate a lower limit $E_{\gamma}$ for our entire sample to determine the fraction of Swift events that could accommodate a jet break beyond the last XRT observation and still be consistent with the relatively narrow pre-Swift $E_{\gamma}$ distribution found by \citet{Frail01} and \citet{Bloom03}.  To analyze the breaks that do exit in a subset of light curves, we employ a Bayesian blocks algorithm \citep{Scargle98, Butler07a} to fit the various segments of XRT light curves, allowing for an automated and robust approach at measuring break times as well as and pre- and post-break power law indices in the afterglow decay. We find that the higher sensitivity of the BAT instrument allows for a large fraction, roughly $\sim 68 \%$, of our sample to accommodate a jet break beyond the last XRT observation and still have an energy consistent with the pre-Swift $E_{\gamma}$ distribution.  We find that relatively few of the X-ray light curves for the remaining events show evidence for breaks in their X-ray light curves that are consistent with that expected from the effects of jet collimation.  The application of these X-ray selected breaks, which typically occur far earlier than the pre-Swift jet breaks observed at optical wavelengths, result in an $E_{\gamma}$ distribution which has a median value that is lower than that of the pre-Swift sample.  We find that the energetics predicted by most of the breaks reported by \citet{Panaitescu07a} suffer from the same difficulties.  The assumed validity of the narrowness of the pre-Swift $E_{\gamma}$ distribution casts doubt on the interpretation of many of these early temporal breaks as jet breaks, unless the intrinsic spread in the collimation corrected energy is much wider than had previously been reported. We discuss the our data acquisition and reduction techniques in $\S 2$ and expand upon our results in $\S 4$.  We discuss the implications of these results $\S 5$.

\section{Data $\&$ Analysis} \label{sec:Data} \label{sec:data}

We form a sample of 76 GRBs detected by Swift with redshifts reported to the Gamma-ray bursts Coordinates Network (GCN) circulars.  The entire list of bursts including redshifts and their associated references can be found on Table 1.  For these, and all other Swift events,  we download the BAT and XRT data products from the {\it Swift}~Archive\footnote{ftp://legacy.gsfc.nasa.gov/swift/data} and process the data with version 0.10.3 of the {\tt xrtpipeline} reduction script and other tools from the HEAsoft~6.0.6\footnote{http://heasarc.gsfc.nasa.gov/docs/software/lheasoft/} software release.  We employ the calibration files from the 2006-10-14 BAT database release for this analysis.  The reduction from cleaned event lists output by the {\tt xrtpipeline} code and from the HEAsoft BAT software to science ready light curves and spectra is described in extensive detail in \citet{Butler07a}.  All of our resulting BAT spectral fits and X-ray light curves to which we apply our analysis are publicly available\footnote{http://astro.berkeley.edu/$\sim$nat/swift}.   The errors regions reported throughout the paper correspond to the 90$\%$ confidence interval.

\subsection{BAT Spectral Fitting} \label{sec:batfitting}

A full and extensive description of the fitting methods use to estimate the bolometric fluence $S_{\rm bol}$ values for our sample of Swift events is discussed in detail in \citet{Butler07b}, although we will briefly summarize our approach here.  We traditionally fit the reduced BAT data for each event with the simplest of three possible models, consisting of a simple powerlaw, a powerlaw times an exponential cutoff, and a smoothly-connected broken powerlaw.  We would then derive confidence intervals by considering random realizations of the data given the best-fit model for each model parameter constrained by the best-fit model. This approach turns out to be very limited for Swift events, due to the narrow energy bandpass of the BAT instrument. In particular it is possible to measure a $\nu F_{\nu}$ spectral peak energy $E_{\rm pk,obs}$ for only about one third of the events in the entire Swift sample. Therefore, for this study we employ a more powerful Bayesian approach which assumes that each burst spectrum has an intrinsic spectrum containing the $E_{\rm pk,obs}$ parameter, and we derive a probability distribution for that parameter given the observed data. We find that the use of prior information, in this case thousands of observations of GRBs by the BATSE instrument \citep{Preece00,Kaneko06}, can be exploited to derive reasonably tight limits on $E_{\rm pk,obs}$ even for cases where $E_{\rm pk,obs}$ is well above the detection passband.  The resulting fits allow us to estimate the true bolometric burst energy fluence despite the limited bandpass of the BAT detector.

To test the validity of our measured $S_{\rm bol}$ and $E_{\rm pk,obs}$ values, we compare them to
values reported to the GCN circulars for 27 events which were detected by both the Swift and Konus-Wind spacecrafts \citep{Aptekar95} and another 7 events detected by both the Swift and Suzaku spacecrafts \citep{Mitsuda07}.  We find that our $S_{\rm bol}$ and $E_{\rm pk,obs}$ values are closely
consistent with the preliminary Konus-Wind and Suzaku values to within a 90$\%$ confidence limit of their reported errors.  Further analysis and comparisons of our $S_{\rm bol}$ and $E_{\rm pk,obs}$ measurements with other instruments for a sample of 216 Swift detected GRBs can be found in \citet{Butler07b}. The final conclusions drawn from that analysis indicates that there is no bias in either the $E_{\rm pk,obs}$ or $S_{\rm bol}$ measurements produced through our Bayesian approach, allowing for a correct bolometric accounting of the total isotropic and/or collimation-correction energy emitted by these events.

\subsection{X-ray Light Curve Region Selection and Fitting} \label{sec:xrayfitting}

\begin{figure}[h] 
\plottwo{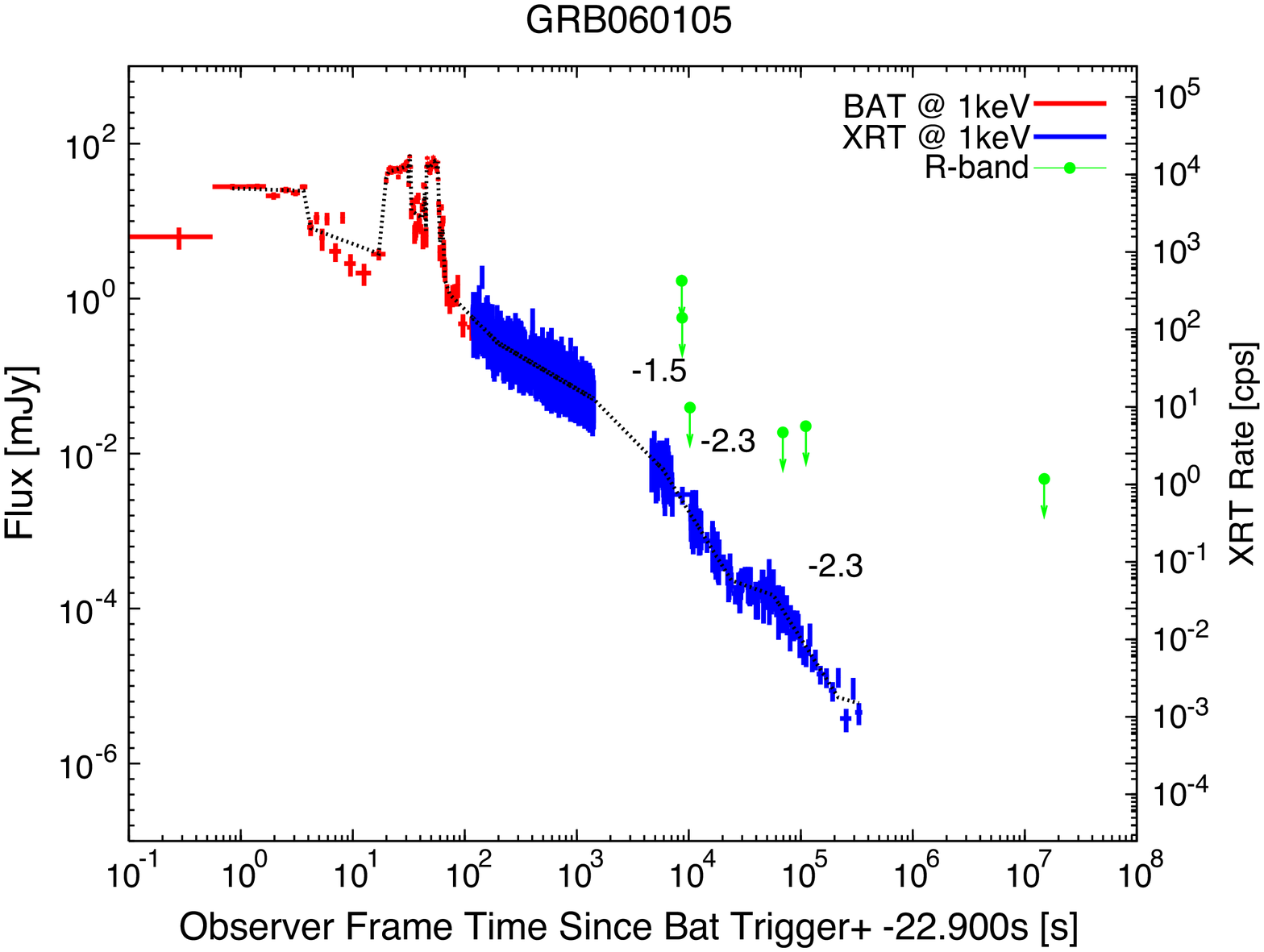}{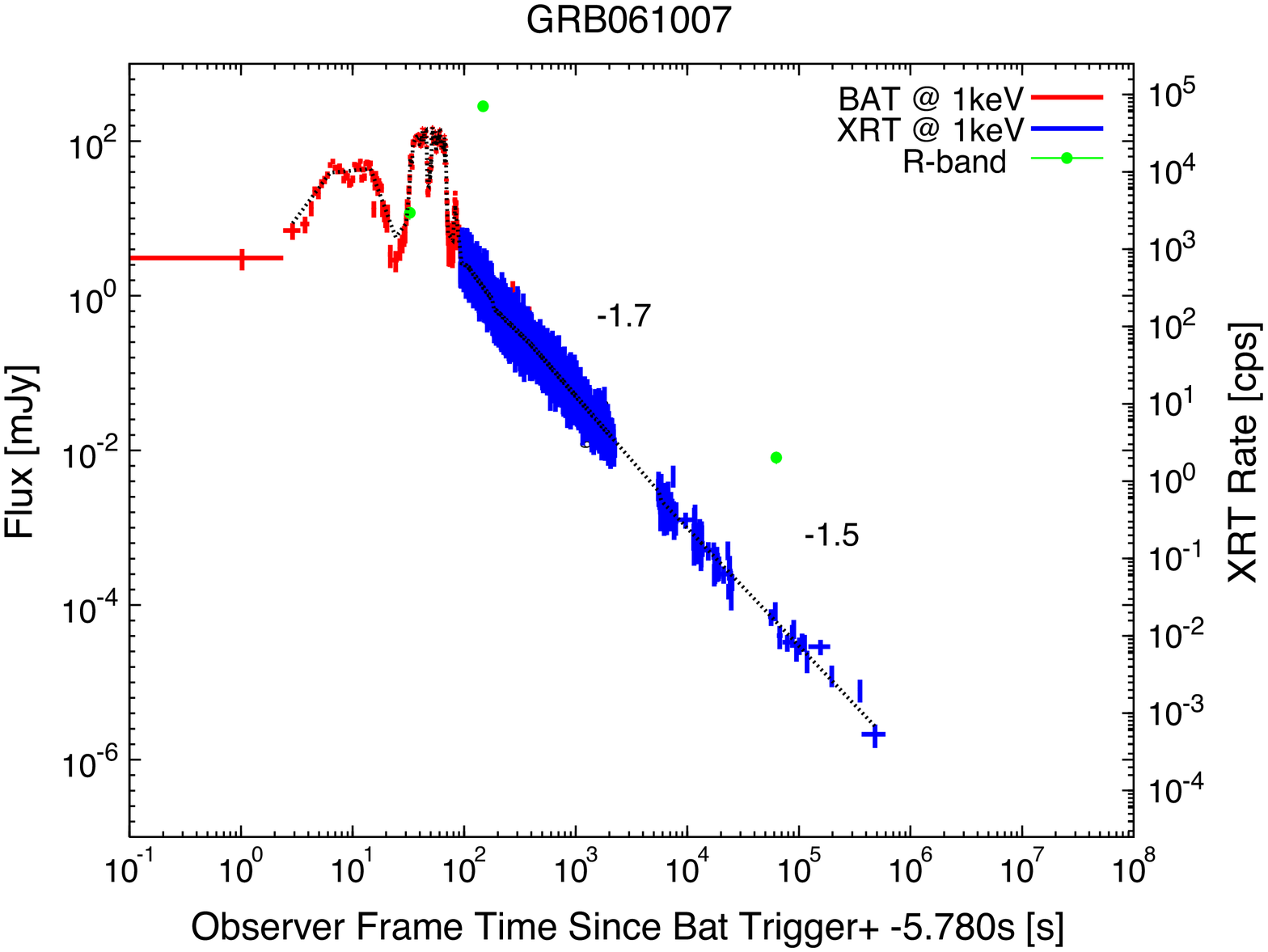}

\caption{Composite BAT (red) and XRT (blue) light curves of four well sampled GRBs exhibiting a characteristic range of potential jet break behavior.  GRB~050820A, GRB~050428A, and GRB~060105 have late-time breaks that are consistent with standard jet models, although GRB~050428A and GRB~060105 have multiple breaks at earlier times that can be misidentified as sudo jet breaks if the last break had not been detected.  The remaining event, GRB~061007, is the quintessential example of a GRB that lacks any breaks in its afterglow light curve. } 
\label{Fig:Lightcurves2}

\end{figure}

In order to measure the temporal power-law indices  $F_{\nu} \propto t^{-\alpha}$ of separate segments in the X-ray light curve, we fit the X-ray light curve data using an extension of the Bayesian blocks algorithm \citep{Scargle98} to piecewise logarithmic data.  Developed in Butler $\&$ Kocevski (2007), the algorithm determines the most likely multi-segment power law fit consistent with the light curves, without the need for human intervention.  The final result of the fitting routine for a sample of events can be seen in Figure~\ref{Fig:Lightcurves2}, where individual light curve segments have been automatically determined and fit to power-laws of various indices\footnote[1]{Similar light curves are available for all Swift bursts at http://astro.berkeley.edu/$\sim$nat/swift}.  For each break, we record the time of occurrence along with the pre- and post-break power law indices, $\alpha_{1}$ and $\alpha_{2}$.    The time of the last 3$\sigma$ detection $t_{\rm XRT}$ of the source by the XRT is also recorded for each event, although this determination does not depend on the Bayesian block algorithm.  The pre- and post-break $\alpha$, the time of the break, and $t_{\rm XRT}$ are listed in Tables 1 $\&$ 2, where available.


\subsection{Isotropic and Collimation Corrected Energy Calculations} \label{sec:Eiso}

We calculate the total isotropic equivalent energy $E_{\rm iso}$ emitted by the GRB from the measured fluence $S_{\rm bol}$ through the standard equation
\begin{equation} \label{eq:eiso}
	E_{\rm iso} = \frac{4 \pi D_{l}^2}{1+z}  S_{\rm bol} k
\end{equation}
where $D_{l}$ is the luminosity distance at redshift $z$ and $k$ represents the multiplicative factor or order unity that translates the bandpass of the detector in the observer frame to a standard rest-frame bandpass, here chosen to be $1-10^{4}$ keV \citep{Bloom01}.  We employ the method outlined in \citet{Amati02} to calculate $k$.  For the case of a homogeneous circumburst medium \citet{Sari99}, the observed jet break time $t_{\rm jet}$ is related to the jet opening angle $\theta_{\rm jet}$ through
\begin{equation} \label{Eq:theta_jet}
	\theta_{jet} = 0.101~{\rm rad}~\left(\frac{t_{\rm jet}}{\rm 1~day}\right)^{3/8} \left(\frac{\xi}{0.2}\right)^{1/8} \left(\frac{n}{10~\rm{cm^{-3}}}\right)^{1/8}\left(\frac{1+z}{2}\right)^{-3/8}\left(\frac{E_{\rm iso}}{10^{53}~{\rm ergs}}\right)^{-1/8}
\end{equation}
where $\xi$ represents the efficiency of converting the blast wave's kinetic energy into gamma-rays and $n$ is the circumburst density.  Throughout our analysis, we have chosen to assume a fixed value for both the efficiency and density parameters, using $\xi=0.5$ and $n=3.0$ \citep{Granot06,Kumar07}.  Although, varying these assumptions within reasonable ranges ($\xi=0.1-1.0$ and $n=3.0-30$) has little effect on our final results.  The conversion from $E_{\rm iso}$ to the collimation corrected energy $E_{\gamma}$ is then a simple geometric correction given by
\begin{equation} \label{Eq:eg}
	E_{\gamma} = 	E_{\rm iso}(1-\cos \theta_{\rm jet})
\end{equation} 
We assume a cosmology with $h = 0.71$, $\Omega_{m} = 0.3$, and $\Omega_{\Lambda} = 0.7$ throughout.

\section{Results} \label{sec:Results}

\begin{figure}[t]
\plotone{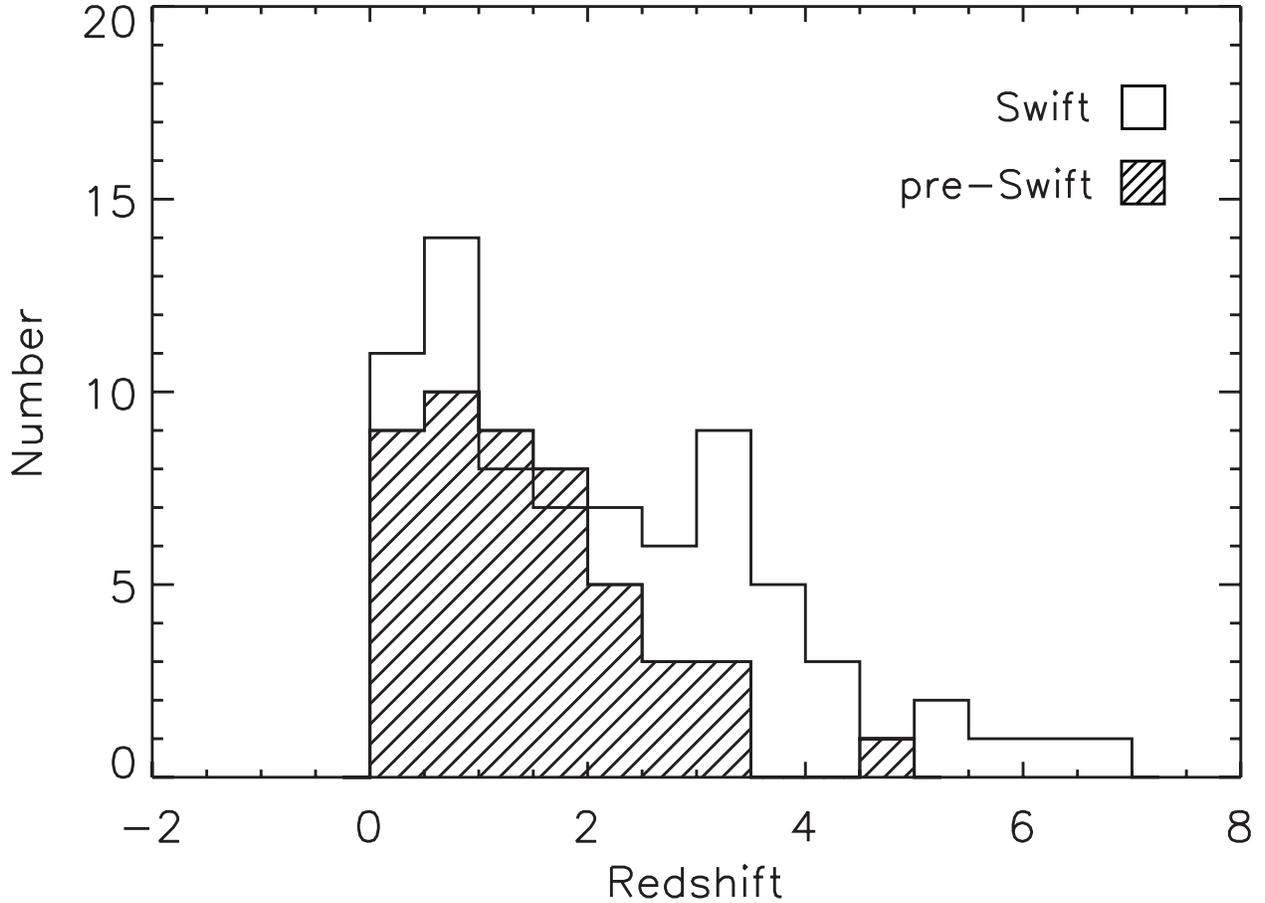}

\caption{The redshift distribution of all 76 GRBs detected by Swift in comparison to the 48 pre-Swift GRBs.  The median redshift of the Swift detected events is slightly higher ($z = 1.8$) than the pre-Swift distribution ($z = 1.1$), owing to the greater sensitivity of the BAT instrument. }

\label{Fig:redshift}
 
\end{figure}

The redshift distribution of all 76 GRBs detected by Swift in comparison to the 48 pre-Swift GRBs \citet{Friedman05} is shown in Figure~\ref{Fig:redshift}.  The median redshift of the Swift detected events is $z = 1.8$ with a standard deviation $\sigma$ of 1.59 compared to the median pre-Swift redshift of $z = 1.1$ and $\sigma$ of 0.98.  A Kolmogorov-Smirnov (K-S) analysis gives the associated probability that the two distributions are consistent to be exceedingly small, at $p = 0.015$. 

For the purposes of comparing our $E_{\rm iso}$, and eventually $E_{\gamma}$, estimates to those of pre-Swift GRBs with known redshift, which consist almost exclusively of long duration events, we form a subset of long bursts (LB) from our original sample of 76 Swift GRBs.  For this set we exclude 9 events which have been classified as short bursts ($t_{\rm 90} \lesssim$ 2 sec), another 3 events which are peculiarly under-luminous (XRF~060218,GRB~GRB~050826,GRB~GRB~051109B) and 1 event (GRB~060124) for insufficient BAT coverage during the prompt emission (see \citet{Butler07b} for more details), leaving a total of 63 LB events.  A plot of $E_{\rm iso}$ vs. redshift for all detected pre-Swift (red) and post-Swift (blue) events (Long, Short, and SN-associated GRBs) is shown in Figure~\ref{Fig:Eiso}.  

The distribution of $E_{\rm iso}$ for the Swift detected GRBs marginally extends to lower energies compared to the pre-Swift sample, even after the admittedly ad hoc exclusion of peculiarly under-luminous events.  The median $E_{\rm iso}$ value of the Swift sample is roughly $4.11~^{+2.53}_{-0.54}\times10^{52}~{\rm erg}$ compared to $7.76~^{+0.01}_{-1.29}\times10^{52}~{\rm erg}$ for the pre-Swift sample.  A K-S test returns a probability of $p = 0.093$.  The inclusion of the under-luminous LB events only worsens the disparity between the two samples.  

How does the observed increase in the median redshift and the decrease in the median $E_{\rm iso}$ of the Swift selected sample effect where we should expected to observe jet breaks in the XRT data?  From Equations \ref{Eq:theta_jet} and  \ref{Eq:eg} we can see that 
\begin{equation} \label{eq:tjet}
t_{\rm jet} \approx 452~\Big(\frac{\xi}{0.2}\Big)^{-1/3}~\Big(\frac{n}{10}\Big)^{-1/3}~\frac{1+z}{2}~\left(\frac{E_{\rm iso}}{10^{53}~{\rm ergs}}\right)^{-1/8}\left[{\rm cos}^{-1}~\Big(1-\frac{E_{\gamma}}{E_{\rm iso}}\Big)\right]^{8/3}~{\rm days}
\end{equation}
Given a fixed $E_{\gamma}$, both effects should increase the observed delay between the initial explosion and the subsequent steepening of the afterglow light curve.  There is sufficient evidence from pre-Swift observations to suggest that $E_{\gamma}$ has a narrow range of values for typical long GRBs.  Using the measured $t_{\rm jet}$, $E_{\rm iso}$, and $z$ for a sample of 17 bursts, \citet{Frail01} and later \citet{Bloom03} concluded that this geometrically corrected energy is narrowly clustered around a standard energy, which \citet{Bloom03} report as $E_{\gamma} = 1.33 \times 10^{51} $~ergs with a variance of 0.35 dex.  Assuming this pre-Swift determined value, and its variance, we can calculate the expected $t_{\rm jet}$ distribution for our Swift detected sample. 

\begin{figure}[t]
\plotone{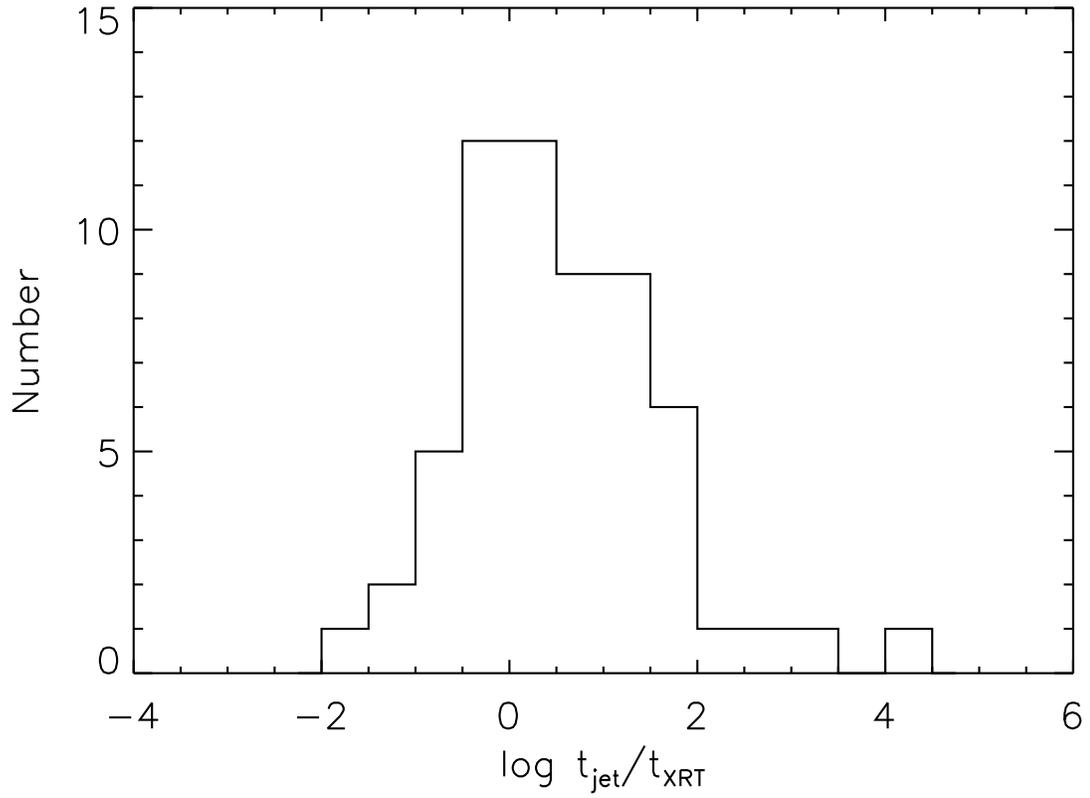}

\caption{A histogram of the ratio between $t_{\rm jet}$ and the last 3$\sigma$ XRT observation $t_{\rm XRT}$.  Roughly $63\% $ of our LB sample have an expected $t_{\rm jet}$, when assuming a fixed pre-Swift $E_{\gamma}$, that occurs beyond the last 3$\sigma$ XRT observation. }
 
 \label{Fig:tjet_expected_ratio}
 
\end{figure}

\begin{figure}[t] \label{Fig:tlast}
\plotone{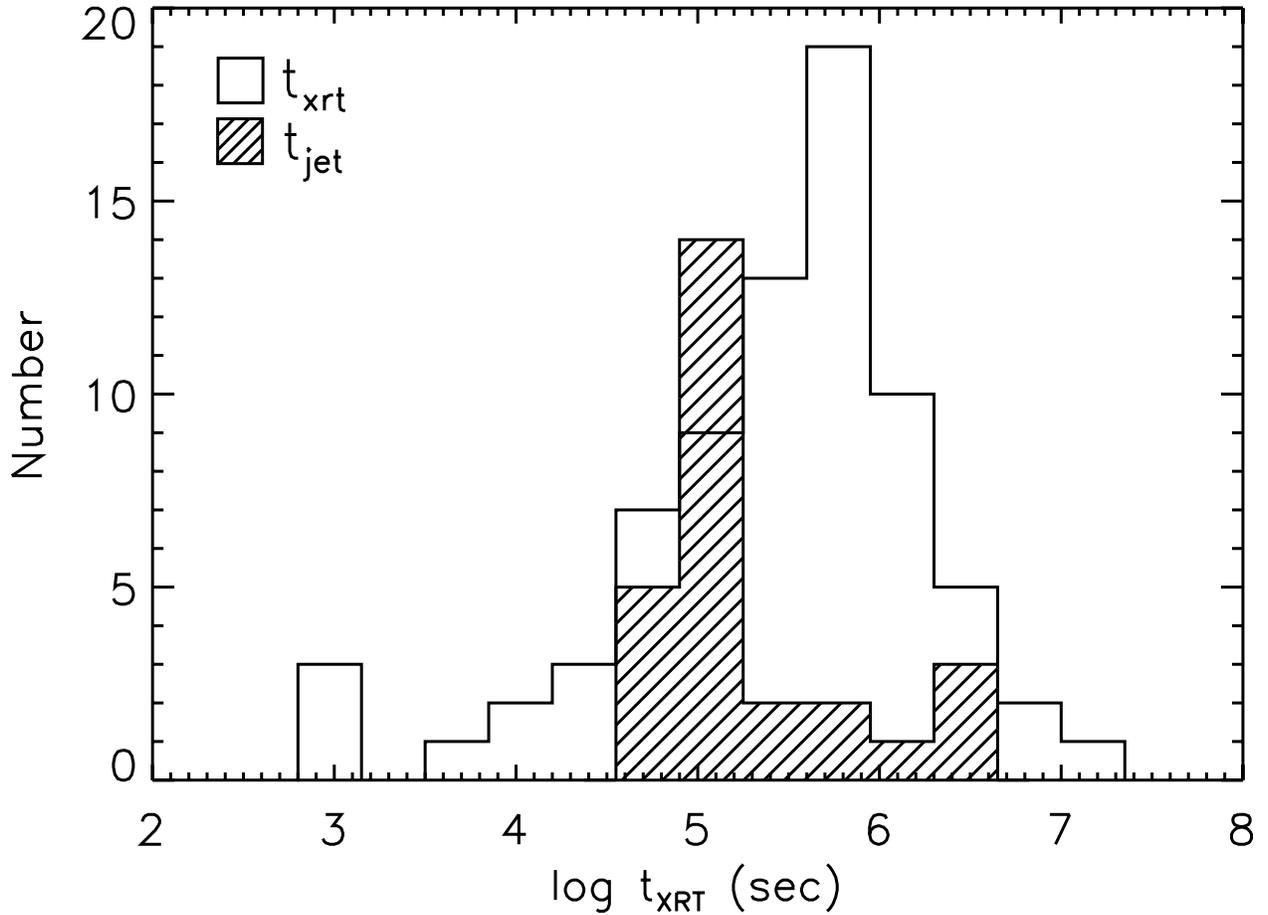}

\caption{A comparison of $t_{\rm XRT}$ for our LB sample to the pre-Swift $t_{\rm jet}$ distribution, with the pre-Swift jet breaks shown as the filled histogram.  The median of the two distributions differ by roughly an order of magnitude.  }

\end{figure}

A histogram of the ratio between this expected $t_{\rm jet}$ and the last 3$\sigma$ XRT observation $t_{\rm XRT}$ is shown Figure~\ref{Fig:tjet_expected_ratio}.  Most events, when assuming a fixed $E_{\gamma}$, have an expected $t_{\rm jet}$ that is near or beyond the last significant XRT detection.  The median expected $\bar{t}_{\rm jet}$ is 10.37 days after the GRB, leading to $63\% $ of our LB sample to have an expected $t_{\rm jet}$ that occurs beyond the last 3$\sigma$ XRT detection.  Three events in our LB subset were excluded from this analysis because their $E_{\rm iso}$ values were below $1.33 \times 10^{51} $~ergs and therefore could not accommodate an $E_{\gamma}$ that was consistent with the pre-Swift energetics distribution.  Such a high ratio of events with $t_{\rm jet} > t_{\rm XRT}$ is somewhat surprising, given that the median duration of XRT observations is typically an order of magnitude longer than the pre-Swift  $t_{\rm jet}$ distribution.  A comparison of $t_{\rm XRT}$ for our LB sample to the pre-Swift $t_{\rm jet}$ is shown in Figure~\ref{Fig:tlast}, with the pre-Swift jet breaks shown as the filled histogram.  The median of the two distributions differ by roughly an order of magnitude.  

What of the light curve breaks that do exist in our LB sample?  Of the 20 GRBs, or $\sim 37\%$ of our sample, for which the expected $t_{\rm jet} < t_{\rm XRT}$, only 3 were determined by \citet{Panaitescu07a} to have jet breaks that are consistent with the standard jet model, 5 events were classified as containing potential jet breaks, and 4 events showed no evidence for late-time breaks in their light curve.  The remaining events in this subset have insufficient coverage to test for the existence of jet breaks.  Of these 8 events that do show some steepening in their light curves, all have breaks that occur earlier than the expected $t_{\rm jet}$, indicating that their $E_{\gamma}$ values must be lower than the assumed standard energy determined from pre-Swift GRBs.  Of the 43 GRBs, or $\sim 63\%$ of our sample, for which the expected $t_{\rm jet} > t_{\rm XRT}$, 8 were determined by \citet{Panaitescu07a} to have jet breaks that are consistent with the standard jet model, 3 events containing potential jet breaks.  For these 11 events with some sign of steepening, their interpretation as jet breaks will necessarily yield $E_{\gamma}$ values that are far less than the narrow peak of the pre-Swift $E_{\gamma}$ distribution because their expected $t_{\rm jet}$ is greater than the last XRT observation.
 
We examine the $E_{\gamma}$ distribution resulting from the use of these early light curve breaks as potential jet breaks, regardless of their predicted $t_{\rm jet}$, to compare this distribution to the pre-Swift $E_{\gamma}$ distributions found by \citet{Frail01} and \citet{Bloom03}.  We do this by utilizing the 25 GRBs in our sample that overlap with the GRBs examined by \citet{Panaitescu07a}.  In this subset, 13 events with redshift were determined to harbor breaks consistent with the standard jet models, and another 12 with ``potential" jet breaks.  Neither the jet break time or the pre- or post-break decay indices were reported by \citet{Panaitescu07a} for the events with ``potential" jet breaks, so we utilize the Bayesian blocks algorithm described in $\S \ref{sec:xrayfitting}$ to measure all three quantities.  The resulting fit parameters are displayed in Table 2. 

\begin{figure}[t] 
\plotone{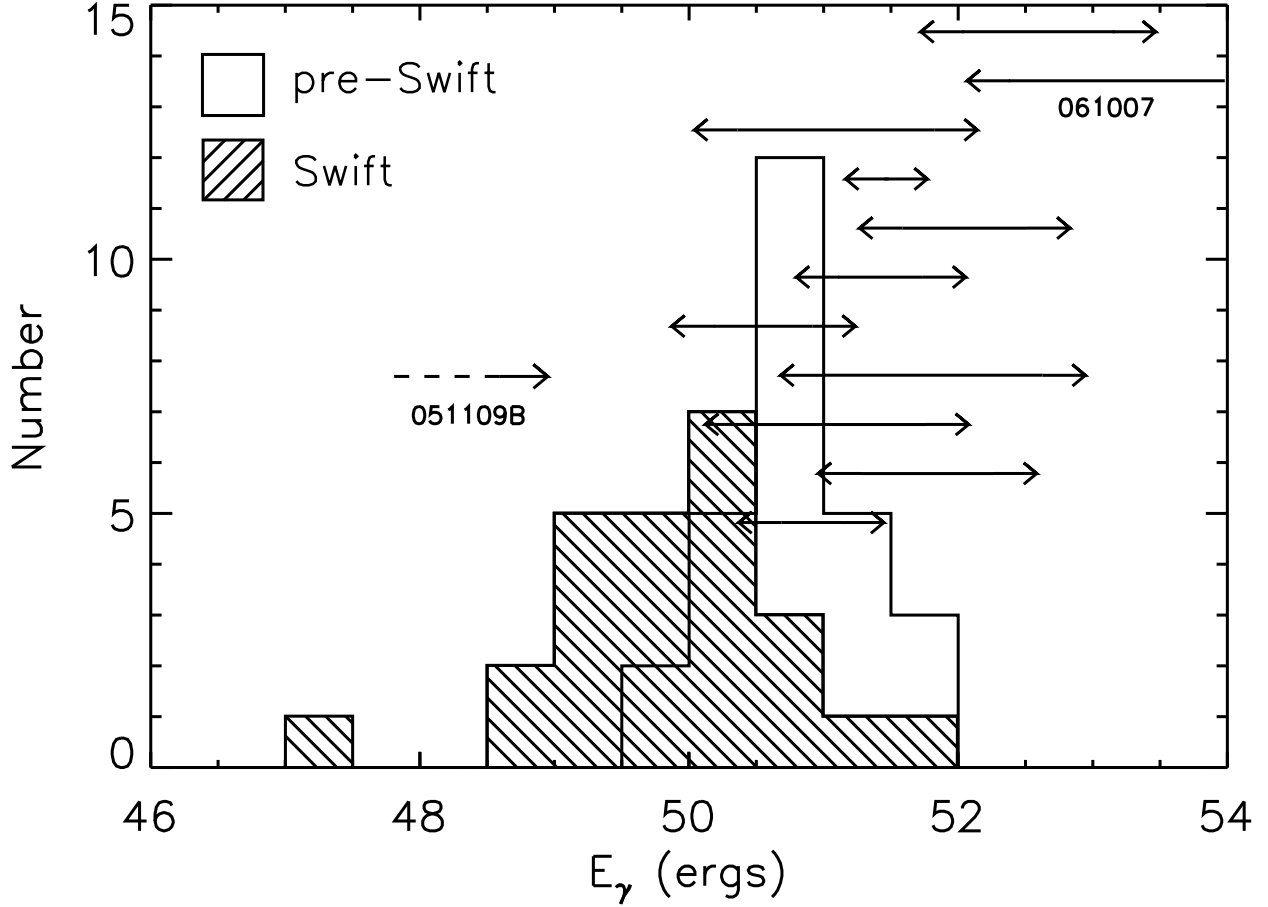}

\caption{A comparison between the pre- and post-Swift collimation corrected energy $E_{\gamma}$ as determined through the use of optical and X-ray determined jet break times, respectively.  The peak of the post-Swift $E_{\gamma}$ distribution is roughly an order of magnitude lower the standard energy derived from the pre-Swift observations.  This is due to Swift's detection of lower energy and higher redshift events as well as the early nature of most X-ray breaks.  Events with no detected breaks in their X-ray light curves have limits to their $E_{\gamma}$ shown as the vertical bars.  The lower limit is set by the last XRT observations whereas the upper limit is the burst's isotropic equivalent energy $E_{\rm iso}$.  Several events begin to push the post-Swift $E_{\gamma}$ distribution beyond $10^{52}$ ergs. }

\label{Fig:eglimitsclass0}

\end{figure}

The resulting post-Swift $E_{\gamma}$ distribution in comparison to the pre-Swift distribution is shown in Figure~\ref{Fig:eglimitsclass0}.  The median of the post-Swift distribution, as determined through the use of all 25 GRBs, is roughly $9.0 \times 10^{49}$~ergs with a variance of 0.90 dex.  The median jet break time for the entire sample is log $t_{\rm jet} \sim 4.73$ days.  The peak of the post-Swift $E_{\gamma}$ distribution is much broader and roughly an order of magnitude lower then the distribution derived from the pre-Swift observations.  A K-S test returns a probability of $p = {5.66 \times 10^{-3}}$ that the pre-Swift and post-Swift distributions are drawn from the same parent population.  The lower and upper limits on $E_{\gamma}$ for all events which show no sign of any breaks in their afterglow light curves are represented as vertical bars in Figure~\ref{Fig:eglimitsclass0}.  The last $3\sigma$ detection of the XRT sets the lower bound whereas the upper bound is set to equal the burst's isotropic equivalent energy $E_{\rm iso}$.  In contrast to the 25 events with breaks in their light curves, the events with no breaks begin to push the post-Swift $E_{\gamma}$ distribution to higher energies in comparison to the pre-Swift distribution.  Two significant outliers (GRB~050820A and GRB~061007) in particular have resulting lower limits to their energies that are above $10^{52}$~ergs.

\section{Discussion} \label{sec:Discussion}


The results presented in the previous section paint a complicated picture for GRB energetics in the Swift era.  The combination of more distant events and a wider distribution in the observed isotropic equivalent energy has resulted in a much broader $E_{\gamma}$ distribution for Swift detected GRBs.  The breaks that do exists in the X-ray light curves of many events are typically inconsistent with standard jet model predictions and occur earlier than the pre-Swift jet break distribution.  Their application as jet breaks yields an under-luminous $E_{\gamma}$ distribution that is highly inconsistent with the pre-Swift sample.  



\begin{figure}[t] 
\plotone{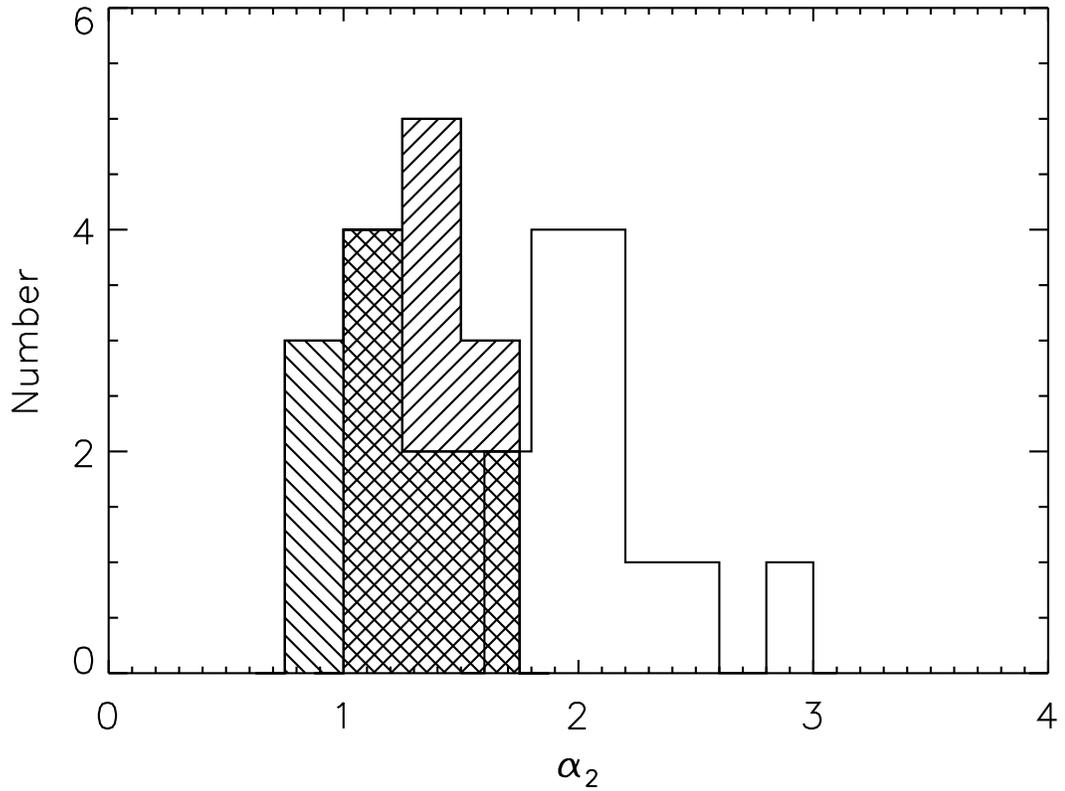}

\caption{A comparison of the post-break decay indices of events containing jet breaks consistent with model predictions (solid histogram) to the post-break decay indices of events containing breaks not fully consistent with jet models (filled histogram at $-45^{\circ}$).  The final decay indices for the events which show no breaks in their light curves is shown as a filled histogram at $+45^{\circ}$. }

\label{Fig:alpha2_comparison}

\end{figure}

It may be the case, as also suggested by \citet{Panaitescu07a}, that a significant fraction of these potential jet breaks are actually associated with some mechanism other than jet collimation, such as the cessation of late-time energy injection \citep{Rees98}.  The events which have post-break decay indices $\alpha_{2}$ that are shallower than the $-$1.5 expected from standard jet models tend to have shallower breaks when compared to the rest of the sample.  The median difference between the pre and post-break decay indices for these bursts is roughly $\Delta\alpha \sim 0.48$ compared to that of bursts that have model consistent jet breaks which are have $\Delta\alpha \sim 1.04$.  Second, the distribution of $\alpha_{2}$ for the events with model inconsistent jet breaks is fully consistent with the decay indices exhibited by events which have no breaks in their light curves, as shown in Figure~\ref{Fig:alpha2_comparison}.  This could indicate that some of these observed breaks may be due to end of the plateau phase that has become ubiquitous in Swift X-ray light curves \citep{Nousek06} and that the true jet break had not been observed by the end of the XRT observations.

\citet{Liang07} has recently completed an extensive analysis of the plateau phases observed in XRT light curves and found that the distribution of transition times to a steeper decay is centered at log $t_{\rm b} = 4.09 \pm 0.61$ seconds and that the distribution of power law indices during the phases peaks at $\alpha_{1} \sim 0.35~\pm~0.35$.  This is in comparison to the median jet break time of log $t_{\rm jet} = 4.72~\pm~0.60$ and a pre-break power law index distribution of $\alpha_{1} \sim 0.82~\pm~0.23$ for model inconsistent jet break in our Swift sample.  Although the distribution of pre-break decay indices is higher than the distribution of plateau decays reported by \citet{Liang07}, it may still be the case that some of the events in our sample consist of light curves with plateau breaks which are steep enough and occur at the upper end of the $t_{\rm b}$ distribution such that they could be considered as jet breaks which do not confirm entirely to standard jet models.  

\begin{figure}[t]
\plotone{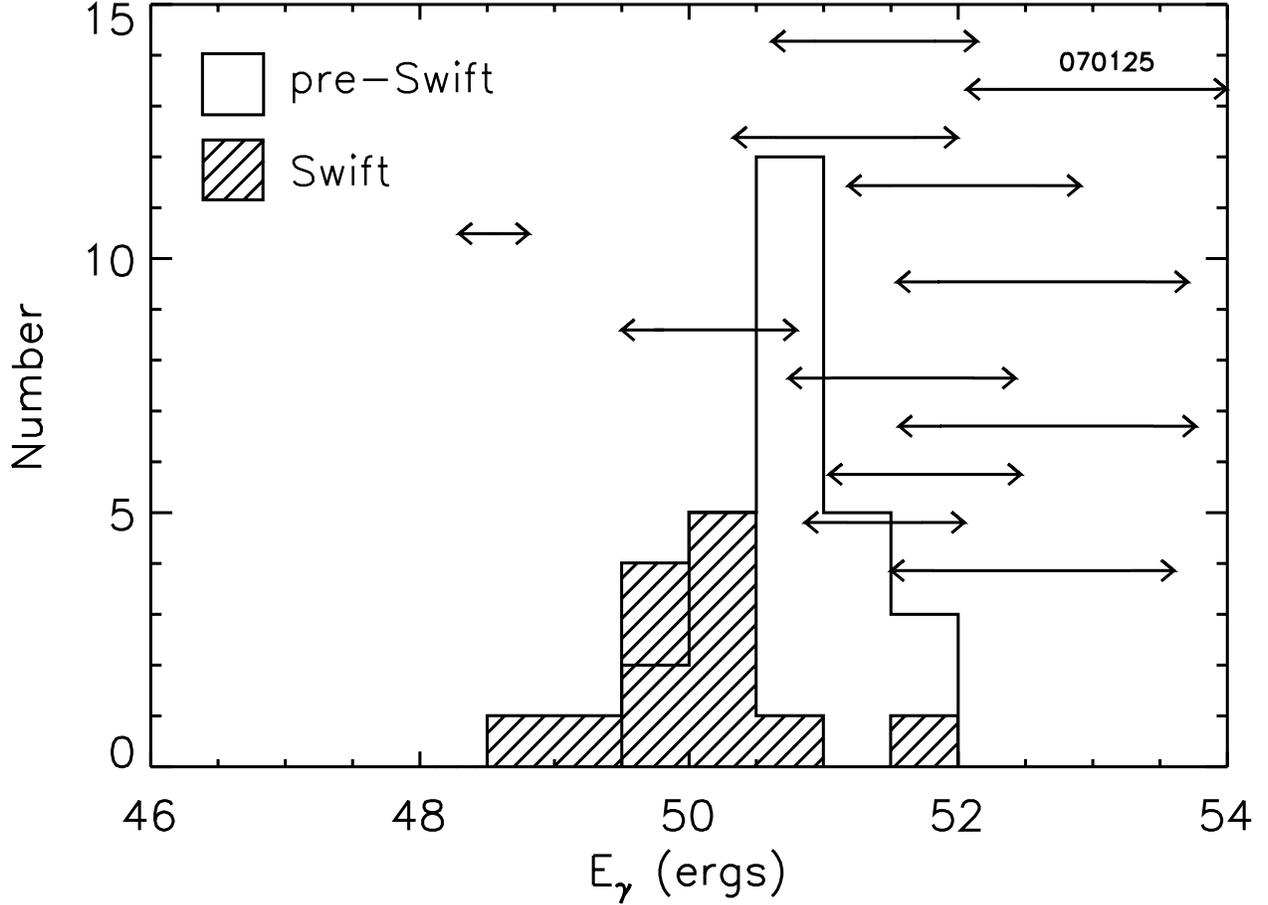}

\caption{Similar to Figure 7,  but the events for which the X-ray breaks have a post-break decay that is shallower than $\alpha \sim 1.5$ have been removed from the sample.  These breaks do not conform with standard jet model predictions and therefore could be considered suspect.  The removal of these events eliminates the low energy tail from the post-Swift $E_{\gamma}$ distribution, making it more consistent with the optically determined pre-Swift energetics distribution.  The upper and lower limits to the $E_{\gamma}$ for these events with shallow breaks are shown as vertical bars.  Much like the events with no detected jet breaks, the limits on $E_{\gamma}$ for many of these shallow break events are well above the pre-Swift $E_{\gamma}$ distribution pushing the upper end of the post-Swift energy distribution well beyond $10^{52}$ ergs.}

 \label{Fig:eglimitsclass0-2}
 
\end{figure}

If we excluding GRBs with X-ray breaks have a post-break decay indices that is shallower than $\alpha \sim 1.5$, we receive a median post-Swift $E_{\gamma} \sim 1.12^{+0.38}_{-2.92}\times 10^{50}$~ergs with a variance of 0.65 dex.  The removal of these events eliminates the low energy tail from the post-Swift $E_{\gamma}$ distribution, making it more consistent with the optically determined pre-Swift energetics distribution, but the distribution is still an order of magnitude lower than the $E_{\gamma}$ distribution estimated from pre-Swift GRBs.  The resulting post-Swift distribution, in comparison to the pre-Swift distribution, is shown in Figure~\ref{Fig:eglimitsclass0-2}.  If we assume that these shallow breaks were due to some mechanism other than jet collimation and that the true jet breaks had not yet manifested by the end of the XRT observations, then we can place upper and lower limits on their collimation corrected energy.  These limits are again represented as vertical bars in Figure~\ref{Fig:eglimitsclass0-2}.  Much like the events with no detected jet breaks, the limits on $E_{\gamma}$ for many of these shallow break events are well above the pre-Swift $E_{\gamma}$ distribution, pushing the upper end of the post-Swift energy distribution well beyond $10^{52}$ ergs.

We note that all of the pre-Swift $E_{\gamma}$ estimates that we have used in this analysis come from jet breaks detected in the optical whereas our sample is limited entirely to X-ray wavelengths.  There is evidence \citep{Panaitescu06,Perley07,Curran07,Oates07} to suggest that the X-ray and optical emission may not evolve achromatically at late times as predicted by blast wave models \citep{Meszaros97,Sari98,Panaitescu00}.  Many of the various light curve phases observed in the X-ray do not always manifest at optical wavelengths, indicating that the X-ray and optical emission may originate from distinct and separate physical components.  Such a two component model has been proposed by \citet{Kumar03} in which the X-ray and optical emissions originate from two distinct jets of differing degrees of collimation.  This model has been invoked by \citet{Panaitescu06} and \citet{Oates07} to explain breaks observed in the X-ray light curves of several events which show no such behavior at optical wavelengths.  In the context of the two component jet model, the X-ray break would be due to a narrow central jet and the optical from a wider jet which presumably would cause a break at a much later time.  Alternatively, \citet{Panaitescu07b} has suggested that the various components in many X-ray light curves may originate from the scattering of forward shock photons by a relativistic outflow behind the leading blast wave.  Under the right conditions, these boosted photons could be more luminous than the X-ray flux of the forward shock itself.  In either case, all of the $E_{\gamma}$ values determined through the use of X-ray breaks become suspect and may not reflect the overall energy budget of the GRB.

These considerations would lead to two possible, albeit it conflicting, scenarios.  First, that the true $E_{\gamma}$ distribution of long GRBs is indeed narrowly clustered about a few times $10^{51}$ ergs as the pre-Swift optical data would suggest but that the breaks seen at X-ray wavelengths are unrelated to jet collimation.  In this case, a majority of the optical breaks would have to occur well beyond $\sim 1$ day.  We find that the median expected jet break time for the Swift sample, given a standard energy of $E_{\gamma} \sim 1.33 \times 10^{51}$ ergs, is roughly $t_{\rm jet} \sim 10$ days, considerably beyond the duration, and indeed capability, of most observing campaigns.  In this case, only 32$\%$ of the Swift sample with known redshift have XRT observations that extend beyond their expected $t_{\rm jet}$.  This scenario would then require some mechanism which could essentially mask the effects of jet collimation at X-ray wavelengths for the events which show straight power law decays to late times.

Even if we dismiss the assumption that the achromaticity of jet breaks extends to X-ray wavelengths, there still appear to be events that challenge the notion of a narrow $E_{\gamma}$ distribution even at optical wavelengths.  The well studied GRB~061007 has an expected jet break at $\sim 0.18$ days but yet exhibits a single power law at optical wavelengths out beyond 1 day, resulting in a lower limit of $E_{\gamma} \sim 1.5 \times 10^{52}$ \citep{Schady06}.  Likewise, 4 events (GRB~050416A, GRB~051016B, GRB~060428B, GRB~060512) have $E_{\rm iso}$ values that are below $10^{51}$ ergs, meaning any break in their optical light curve would make the inferred energy released by these events even more inconsistent with pre-Swift estimates. 

This leads to the second scenario in which the jet breaks are indeed seen in some X-ray light curves but that long cosmological GRBs have a much wider distribution of $E_{\gamma}$ then pre-Swift GRBs would suggest.  A wider $E_{\gamma}$ distribution has already been suggested by low energy events such as XRFs and SN-GRBs and there is no a priori reason to believe that the gamma-rays, which represent such a sub-dominate fraction of the entire energy in the collapsar model, should necessarily be standard among all events.  It could then still be the case that some the breaks seen in the X-ray wavelengths could be due to jet collimation while others may be manifestation of other mechanisms that only effect the resulting X-ray emission.  

\begin{figure}[t] 
\plotone{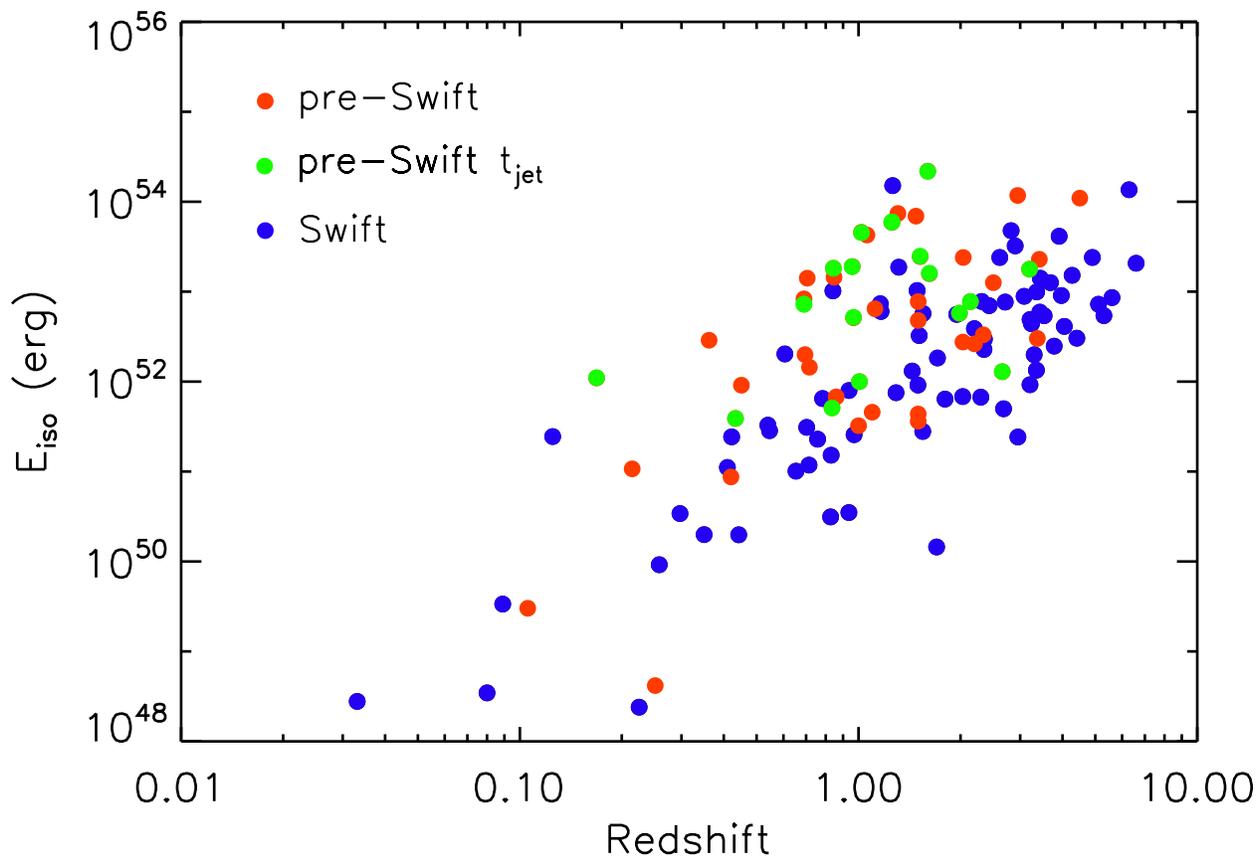}

\caption{A plot of $E_{\rm iso}$ vs. redshift for all detected pre-Swift (red) and post-Swift (blue) events (Long, Short, and SN-associated GRBs).  The pre-Swift events for which a jet break was detected are shown in green.  These pre-Swift GRBs that contain jet break tend to populate a high $E_{\rm iso}$ and intermediate redshift parameter space.  Overall, the Swift detected distribution of $E_{\rm iso}$ extends to lower energies compared to the pre-Swift sample.  }

\label{Fig:Eiso-z}

\end{figure}

In this case, the disparity between the pre- and post-Swift energetics distributions would imply a significant bias towards the detection of jet breaks in bright, relatively nearby, GRBs in the pre-Swift era.  The effects of such a bias begin to be apparent in Figure~\ref{Fig:Eiso-z}, where $E_{\rm iso}$ is plotted vs. redshift for all pre- (red) and post- (blue) Swift events, including long, short, and SN-associated GRBs.  The events for which jet breaks were detected in the pre-Swift sample are plotted in green and represent the brighter fraction of the distribution.  Furthermore, the presence of multiple light curve breaks and the long delay between the GRB and subsequent optical and X-ray observations prior to the Swift mission may have also allowed for a bias towards the selection of the last break in a multi-break light curve as the potential jet break.  This would have essentially created an artificial lower limit on the pre-Swift $E_{\gamma}$ distribution.
 
Ultimately, more contemporaneous observations at optical and X-ray wavelengths will be needed to determine which fraction of the X-ray breaks are truly achromatic.  Observing more jet breaks at optical wavelengths would also allow for a direct comparison between the pre- and post-Swift energetics distribution, hopefully eliminating the ambiguity involved with using jet break times determined at different wavelengths.  This will require a continued interest by the GRB community in obtaining deep optical and/or IR imaging of afterglows several days or even weeks after the initial explosion.  Such observations will prove crucial in resolving many of these questions regarding the collimation and energetics of GRBs.

\section{Acknowledgments} \label{sec:acknowledgments}

D.K. acknowledges financial supported through the NSF Astronomy $\&$ Astrophysics Postdoctoral Fellowships under award AST-0502502. N.B. gratefully acknowledges support from a Townes Fellowship at U. C. Berkeley Space Sciences Laboratory and partial support from J. Bloom and A. Filippenko.






\bigskip 

\section*{Figures}

\clearpage

\begin{figure}[t] 
\plottwo{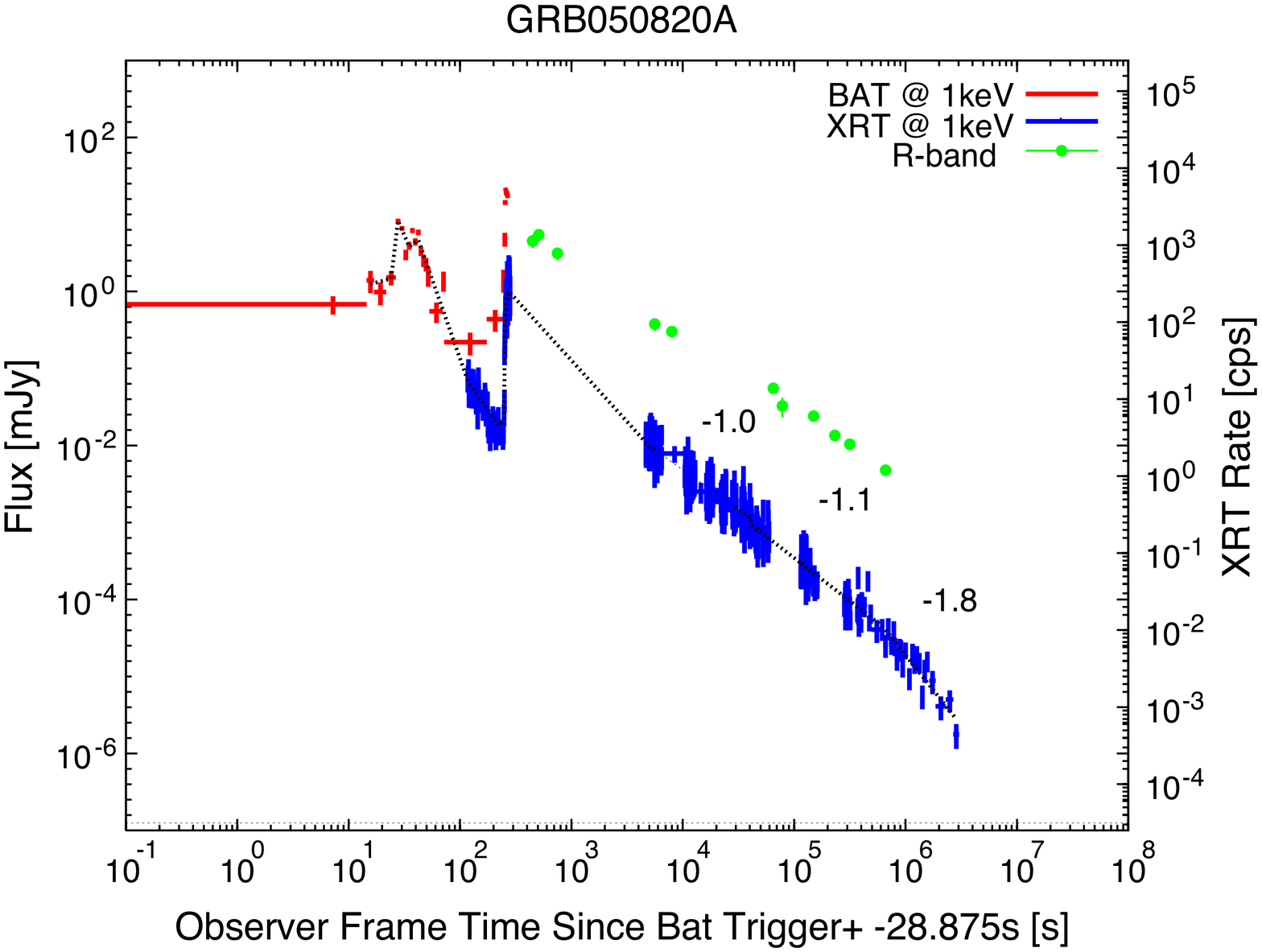}{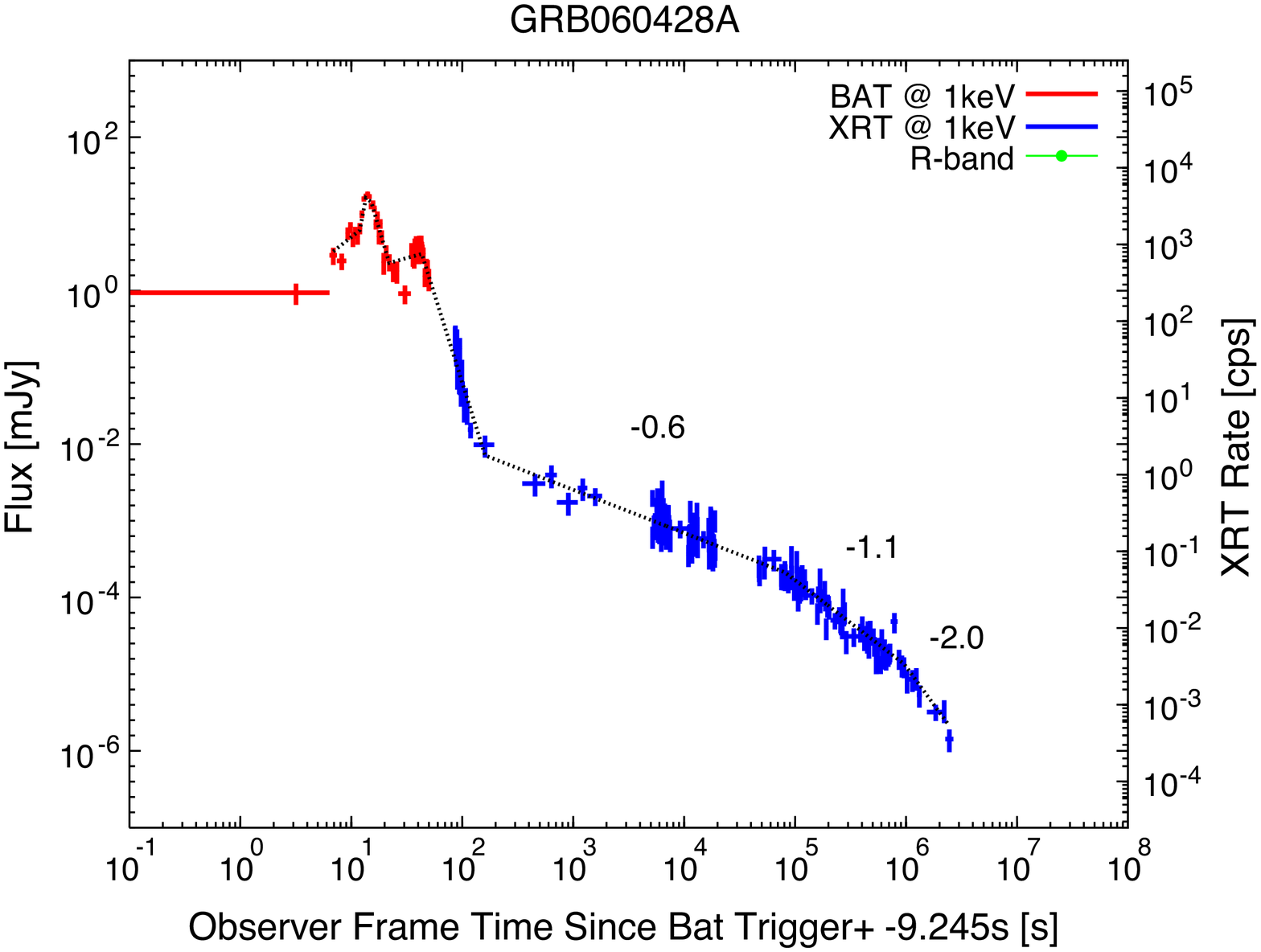}
\label{Fig:Lightcurves1}
\end{figure}

\clearpage

\clearpage



\begin{figure} 
\plotone{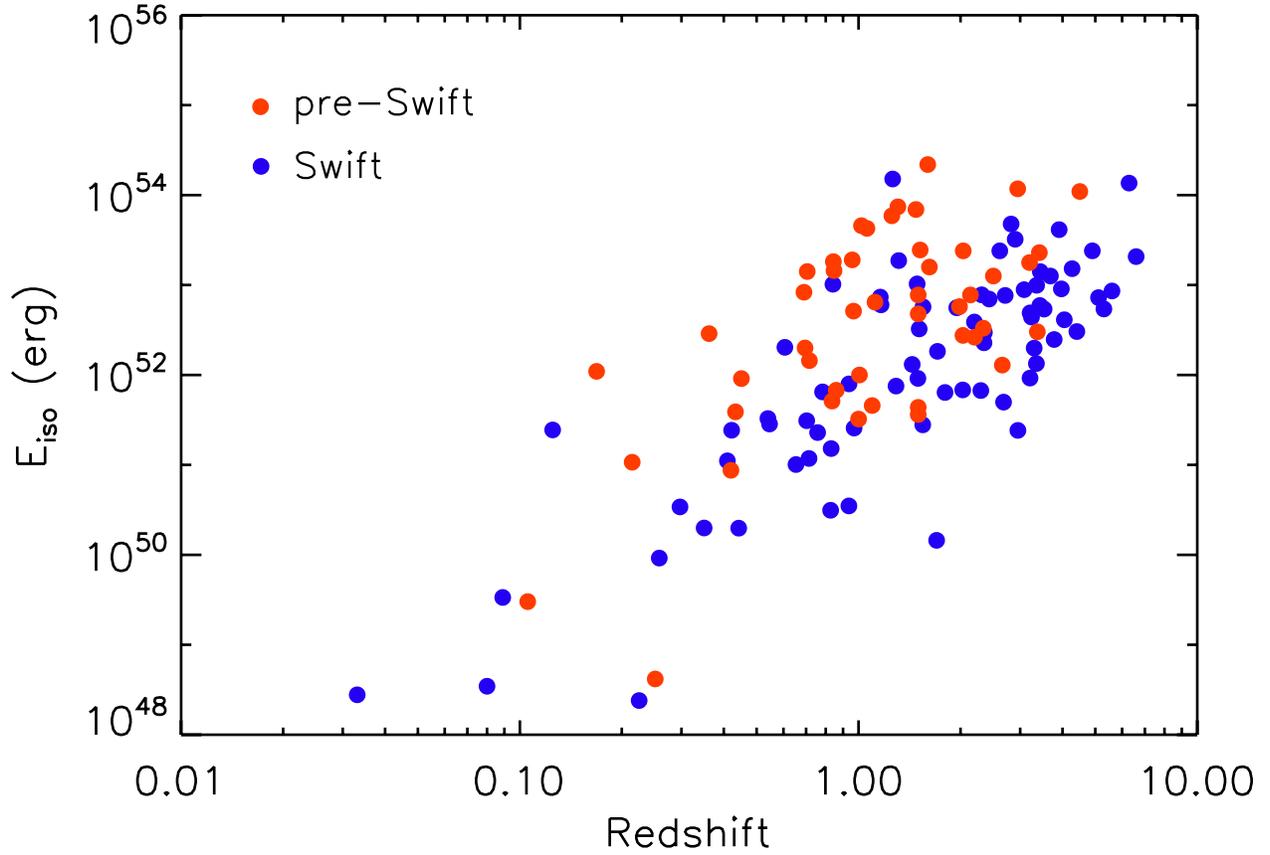}

\caption{A plot of $E_{\rm iso}$ vs. redshift for all detected pre-Swift (red) and post-Swift (blue) events (Long, Short, and SN-associated GRBs).  Overall, the Swift detected distribution of $E_{\rm iso}$ extends to lower energies compared to the pre-Swift sample.  The apparent correlation is a result of an unknown population evolution combined with detector threshold effects  (see Butler, Kocevski, et al. 2007).  }
\end{figure}

\label{Fig:Eiso}

\clearpage

\clearpage

\clearpage

\clearpage

\clearpage

\clearpage

\clearpage






\begin{deluxetable}{lrrrrrrrr}

\label{Table:sample} \tablecolumns{5} \tablewidth{0pc}
\tablecaption{76 Swift Detected GRBs With Measured Redshift}
\tablehead{ \colhead{GRB\tablenotemark{a}} & \colhead{Redshift\tablenotemark{b}} & \colhead{log $E_{\rm iso}$}
& \colhead{log $t_{\rm XRT}$\tablenotemark{c}} & \colhead{log $E_{\gamma,\rm XRT}$\tablenotemark{d}}
\\
\colhead{} & \colhead{} & \colhead{(ergs)}  & \colhead{(s)} & \colhead{(ergs)}  }

\startdata

050126	  &        1.29$^*$  &         51.88 (0.13,0.38) 	     &  4.77	   &     $>$51.90	 \\
050315     &        1.95  &         52.75 (0.00,0.33)       &    5.98	  &  	 $>$53.05 \\
050318     &        1.44$^*$  &         52.12 (0.15,0.02)       &    4.80	  &  	 $>$51.88 \\
050319     &        3.24  &         52.64 (0.04,0.40)       &    4.97	  &  	 $>$51.25 \\
050401     &        2.90  &         53.51 (0.10,0.25)       &    5.88	  &  	 $>$53.67 \\
050416A     &       0.65$^*$   &        51.00 (0.11,0.17)        &   6.81	   & 	 $>$51.23 \\
050505     &        4.27  &         53.18 (0.10,0.28)       &    5.70	  &  	 $>$53.45 \\
050509B$\dagger$     &       0.22   &        48.38 (0.22,0.46)        &  	\nodata	   & 	\nodata	    \\
050525     &        0.61  &         52.31 (0.02,0.02)       &    6.05	  &  	 $>$51.99 \\
050603     &        2.82  &         53.68 (0.21,0.27)      &     5.78 	 &   	 $>$53.98 \\
050724$\dagger$     &        0.26  &         49.96 (0.10,0.33)      &     5.26 	 &    	 $>$49.67 \\
050730     &        3.97  &         52.96 (0.16,0.28)      &     5.68 	 &   	 $>$53.26 \\
050802     &        1.71  &         52.26 (0.09,0.27)      &     5.90 	 &   	 $>$52.11 \\
050803     &        0.42$^*$  &         51.39 (0.18,0.30)      &     5.92 	 &   	 $>$51.51 \\
050813$\dagger$     &        1.70  &         50.16 (0.34,0.44)      &     3.04 	 &   	 $>$50.44 \\
050814     &        5.30$\ddagger$     &         52.73 (0.11,0.18)      &     5.94 	 &   	 $>$52.39 \\
050820A     &       2.61   &        53.38 (0.26,0.25)       &    6.70 	  &  	 $>$53.58 \\
050824     &        0.83  &         51.18 (0.15,0.82)      &     6.31 	 &   	 $>$51.30 \\
050826     &        0.30$^*$  &         50.53 (0.36,0.33)      &     5.40	 &   	 $>$49.37 \\		
050904     &        6.29  &         54.13 (0.17,0.17)      &     4.73	 &   	 $>$54.40 \\ 
050908     &        3.35  &         52.13 (0.12,0.21)      &     4.97	 &   	 $>$51.35 \\ 
050922C     &       2.20   &        52.59 (0.09,0.23)       &    5.00	  &  	 $>$51.87 \\ 
051016B     &       0.94$^*$   &        50.54 (0.06,0.42)       &    5.73	  &  	 $>$50.85 \\ 
051109A     &       2.35   &        52.36 (0.11,0.31)       &    6.19	  &  	 $>$52.33 \\ 
051109B     &       0.08$^*$   &        48.54 (0.11,0.20)       &    4.92	  &  	 $>$46.87 \\ 
051111     &        1.55  &         52.76 (0.12,0.28)      &     4.67	 &   	 $>$52.33 \\ 
051221A$\dagger$     &       0.55   &        51.45 (0.19,0.24)       &    5.86	  &  	 $>$51.33 \\ 
051227$\dagger$    &        0.71$^*$  &         51.07 (0.26,0.36)      &     5.21 	 &   	 $>$51.34 \\ 
060108     &        2.03$\ddagger$   &         51.83 (0.13,0.32)      &     5.53 	 &   	 $>$51.73 \\ 
060115     &        3.53  &         52.73 (0.03,0.19)      &     5.67 	 &   	 $>$52.92 \\ 
060116     &        6.60$\ddagger$     &         53.32 (0.16,0.26)      &     5.20 	 &   	 $>$53.27 \\ 
060124     &        2.30  &         51.83 (0.11,0.31)      &     6.28 	 &   	 $>$51.07 \\ 
060202     &        0.78$^*$  &         51.81 (0.08,0.29)      &     6.44 	 &   	 $>$50.62 \\ 
060206     &        4.05  &         52.61 (0.07,0.11)      &     6.57 	 &   	 $>$52.89 \\ 
060210     &        3.91  &         53.62 (0.09,0.27)      &     5.94 	 &   	 $>$53.90 \\ 
060218     &        0.03  &         48.44 (0.10,0.18)      &     5.90 	 &   	 $>$47.66 \\ 
060223A     &       4.41   &        52.48 (0.07,0.15)       &    4.49	  &  	 $>$52.74 \\ 
060418     &        1.49  &         53.01 (0.10,0.23)      &     5.52	 &   	 $>$53.17 \\ 
060428B     &       0.35$^*$   &        50.30 (0.10,0.29)       &    5.98	  &  	 $>$49.81 \\ 
060502A     &       1.51   &        52.51 (0.14,0.27)       &    6.20	  &  	 $>$52.81 \\ 
060502B$\dagger$     &       0.09$^*$   &        49.53 (0.33,0.41)       &    3.15	  &  	 $>$49.35 \\ 
060510B     &       4.90   &        53.38 (0.10,0.14)       &    5.16	  &  	 $>$53.21 \\ 
060512     &        0.44$^*$  &         50.30 (0.10,0.40)      &     5.39	 &   	 $>$47.96 \\ 
060522     &        5.11  &         52.86 (0.09,0.29)      &     5.38	 &   	 $>$52.20 \\ 
060526     &        3.21  &         52.69 (0.01,0.34)      &     5.33	 &   	 $>$52.79 \\ 
060604     &        2.68  &         51.70 (0.09,0.53)      &     5.90 	 &   	 $>$50.66 \\ 
060605     &        3.78  &         52.39 (0.13,0.35)      &     4.84 	 &   	 $>$51.85 \\ 
060607A     &       3.08   &        52.95 (0.08,0.25)       &    5.30 	  &  	 $>$52.68 \\ 
060614$\dagger$     &        0.12  &         51.39 (0.09,0.06)      &     5.85 	 &   	 $>$51.17 \\ 
060707     &        3.43  &         52.77 (0.06,0.13)      &     6.23 	 &   	 $>$52.95 \\ 
060708     &        1.80$\ddagger$     &         51.81 (0.09,0.19)      &     5.80 	 &   	 $>$51.18 \\ 
060714     &        2.71  &         52.88 (0.05,0.30)      &     5.81 	 &   	 $>$51.82 \\ 
060729     &        0.54  &         51.52 (0.08,0.28)      &     7.12 	 &   	 $>$51.60 \\ 
060904B     &       0.70   &        51.49 (0.11,0.20)       &    5.27 	  &  	 $>$51.76 \\ 
060906     &        3.68  &         53.10 (0.04,0.30)      &     4.85	 &   	 $>$53.17 \\ 
060908     &        2.43  &         52.84 (0.09,0.19)      &     5.90	 &   	 $>$52.53 \\ 
060912A     &       0.94$^*$   &        51.90 (0.14,0.21)       &    5.39	  &  	 $>$51.95 \\ 
060926     &        3.21  &         51.97 (0.09,0.55)      &     5.06	 &   	 $>$51.99 \\ 
060927     &        5.47  &         52.93 (0.07,0.09)      &     3.79	 &   	 $>$52.98 \\ 
061004     &        3.30  &         52.30 (0.06,0.20)      &     5.17	 &   	 $>$52.26 \\ 
061007     &        1.26  &         54.18 (0.24,0.22)      &     5.78	 &   	 $>$54.47 \\ 
061028     &        0.76$^*$  &         51.36 (0.12,0.36)      &     4.39	 &   	 $>$51.01 \\ 
061110B     &       3.44   &        53.15 (0.36,0.31)       &    4.21	  &  	 $>$53.42 \\ 
061121     &        1.31  &         53.27 (0.11,0.21)      &     6.34 	 &   	 $>$53.31 \\ 
061126     &        1.16$^*$  &         52.87 (0.15,0.29)      &     6.13 	 &   	 $>$52.38 \\ 
061210$\dagger$     &        0.41$^*$  &         51.05 (0.43,0.35)      &     5.58 	 &   	 $>$50.54 \\ 
061217$\dagger$     &        0.83$^*$  &         50.50 (0.42,0.34)      &     4.12 	 &   	 $>$50.18 \\ 
061222B     &       3.36   &        53.00 (0.18,0.19)       &    4.77 	  &  	 $>$53.08 \\ 
070110     &        2.25  &         52.47 (0.09,0.27)      &     6.36 	 &   	 $>$52.69 \\ 
070208     &        1.17  &         51.45 (0.14,0.26)      &     5.30 	 &   	 $>$51.09 \\ 
070306     &        1.50$^*$  &         52.78 (0.08,0.28)      &     6.06 	 &   	 $>$50.55 \\ 
070318     &        0.84  &         51.96 (0.12,0.29)      &     6.00 	 &   	 $>$52.25 \\ 
070411     &        2.95  &         53.01 (0.09,0.25)      &     5.84 	 &   	 $>$53.31 \\ 
070419A     &       0.97   &        51.38 (0.11,0.28)       &    3.07 	  &  	 $>$51.44 \\ 
070506     &        2.31  &         51.41 (0.10,0.22)      &     3.91 	 &   	 $>$51.47 \\ 
070508     &        0.82$^*$  &         52.89 (0.07,0.08)      &      5.51  &   	 $>$53.18 \\ 
\enddata

\tablenotetext{a} {Short GRBs are denoted by a dagger and are excluded from our primary analysis.}
\tablenotetext{b} {Photometrically determined redshifts are marked with a double dagger.  Redshifts determined through host association are denoted with an asterisk.  A full reference list for all quoted redshifts can be found in \citet{Butler07b}. }
\tablenotetext{c} {The last 3$\sigma$ detection of the afterglow by the XRT.}
\tablenotetext{d} {The lower limit on $E_{\gamma}$ when using the last XRT observation as the lower limit to the jet break time.}

\end{deluxetable}

\begin{deluxetable}{lrrrrrrrr}

\label{Table:JetBreaks} \tablecolumns{5} \tablewidth{0pc}
\tablecaption{GRBs With Potential Jet Breaks}
\tablehead{ \colhead{GRB\tablenotemark{a}} & \colhead{Redshift} & \colhead{log $t_{\rm jet}$} & \colhead{$\alpha_{1}$} & \colhead{$\alpha_{2}$} & \colhead{log $E_{\gamma}$}
\\
\colhead{} & \colhead{} & \colhead{(s)}  & \colhead{}  & \colhead{} & \colhead{(ergs)}  }

\startdata

050315	  &       1.95	 &  5.41	  &    0.64	  &  	1.91   &  	50.72	  \\
050318    &       1.44   &    4.41   &     1.35  &      2.06  &     49.56    \\
050319    &       3.24   &    4.78   &     0.67  &      1.71  &     50.05    \\
050505    &       4.27   &    4.84   &     1.27  &      1.90  &     50.43    \\
050730    &      3.97    &   4.11    &    0.82   &     2.57   &    49.73     \\
050803    &       0.42   &    4.11   &     0.29  &      1.76  &     48.96    \\
050814    &      5.30    &   4.94    &    0.81   &     2.24   &    50.11     \\
051022    &       0.80   &    5.41   &     1.40  &      2.18  &     51.54    \\
060526    &       3.21   &    5.11   &     0.89  &      2.91  &     50.34    \\
060605    &       3.78   &    4.11   &     1.04  &      2.10  &     49.33    \\
060614    &       0.12   &    5.11   &     1.29  &      2.14  &     49.79    \\
060906    &       3.68   &    4.24   &     0.43  &      1.81  &     49.96    \\
070306    &       1.50   &    4.78   &     1.20  &      1.91  &     50.33    \\
\nodata & \nodata & \nodata & \nodata & \nodata & \nodata \\
050401	  &   	  2.90	 &    3.39	 & 		0.66  &  	1.38   &  	49.69	\\
050408    &      1.24    &   4.55    &    0.70   &     1.00   &    49.50     \\
050525A   &        0.61  &    3.78   &     0.91  &      1.39  &     49.37    \\
050603    &       2.82   &    4.83   &     1.40  &      1.70  &     50.90    \\
050802    &       1.71   &    3.81   &     0.73  &      1.25  &     49.18    \\
051016B   &        0.94  &    4.80   &     0.80  &      1.17  &     48.74    \\
060210    &      3.91    &   4.67    &    0.98   &     1.28   &    50.65     \\
060218    &      0.03    &   4.73    &    0.82   &     1.23   &    47.32     \\
060707    &     3.43     &  4.72     &   0.60    &    1.05    &   50.09      \\
060708    &    1.80      & 4.20      &  0.74     &   1.38     &  49.13       \\
070125    &    1.55      & 5.11      &  0.90     &   1.60     &  51.49       \\
070318    &    0.84      & 5.44      &  1.17     &   1.71     &  50.31       \\
\enddata

\tablenotetext{a} {The first group consists of GRBs with spectral and temporal behavior found by \citet{Panaitescu07a} to be consistent with standard jet models.  The second group have light curve breaks that are not fully consistent with model predictions, but have exhibit steepening that resembles jet break behavior.}

\end{deluxetable}



\begin{thebibliography}{}

\bibitem[Amati et al.(2002)]{Amati02} Amati, L., \aap, 390, 81
\bibitem[Aptekar et al.(1995)]{Aptekar95} Aptekar et al. 1995, SSRv, 71, 265A
\bibitem[Barthelmy et al.(2005b)]{Barthelmy05a} Barthelmy, S.~D., et al. 2005, Space Science Reviews, 120, 143
\bibitem[Barthelmy et al.(2005a)]{Barthelmy05b} Barthelmy, S. D., et al. 2005a, 2005, \apj, 635L, 133
\bibitem[Bloom, Frail, $\&$ Sari(2001)]{Bloom01} Bloom, J., S., Frail, D., A., Sari, R. 2001, AJ, 121.2879B
\bibitem[Bloom, Frail, $\&$ Kulkarni(2003)]{Bloom03} Bloom, J. S.; Frail, D. A.; Kulkarni, S. R 2003, ApJ, 594, 674B
\bibitem[Burrows et al.(2005a)]{Burrows05a} Burrows, D.~N., et al. 2005, Space Science Reviews, 120, 165
\bibitem[Burrows et al.(2005b)]{Burrows05b} Burrows, D. N., et al. 2005, Science, 309, 1833
\bibitem[Burrows et al.(2007)]{Burrows07} Burrows, D. N., et al. 2007, Submitted to Philosophical Transactions (astro-ph/0701046)
\bibitem[Butler $\&$ Kocevski(2007)]{Butler07a} Butler, N. $\&$ Kocevski, D. 2007, ApJ, 663, 407B
\bibitem[Butler et al.(2007)]{Butler07b} Butler, N., Kocevski, D., Bloom, J. S., Curtis, J. L. 2007, Submitted to \apj (2007arXiv0706.1275B) 
\bibitem[Curran et al.(2007)]{Curran07} Curran, P, .A., et al. 2007, Submitted to MNRAS Letters (arXiv0706.1188C) 
\bibitem[Frail et al.(2001)]{Frail01} Frail, D. A., et al. 2001, ApJ, 562, L55 
\bibitem[Friedman $\&$ Bloom(2005)]{Friedman05} Friedman, A., S.; Bloom, J., S. 2005, ApJ, 627, 1F
\bibitem[Frontera et al.(2000)]{Frontera00} Frontera, F., et al. 2000, \apj Suppl., 127, 59-78
\bibitem[Fruchter et al.(1999)]{Fruchter99} Fruchter, A. S., et al. 1999, ApJ, 519, L13
\bibitem[Gehrels et al.(2004)]{Gehrels04} Gehrels, N., et al. 2004, \apj, 611, 1005 
\bibitem[Gendre et al.(2006)]{Gendre06} Gendre, B., 2006, A$\&$A 455, 803
\bibitem[Ghirlanda, Ghisellini, $\&$ Lazzati(2004)]{Ghirlanda04} Ghirlanda, G., Ghisellini, G., $\&$ Lazzati D. 2004, ApJ, 616, 331
\bibitem[Granot, Kšnigl, $\&$ Piran(2006)]{Granot06} Granot, J., Kšnigl A. $\&$ Piran, T. 2006, MNRAS, 370, 1946
\bibitem[Harrison et al.(1999)]{Harrison99} Harrison, F. A., et al. 1999, ApJ, 523, L121 
\bibitem[Holland et al.(2007)]{Holland07} Holland et al. 2007, AJ, 133, 122H
\bibitem[Kaneko et al.(2006)]{Kaneko06} Kaneko, Y., Preece, R., Briggs, M., S., Paciesas, W., S., Meegan, C., A.; Band, D., L. 2006, ApJS, 166, 2
\bibitem[Kolaczyk \& Dixon(2000)]{Kol00} Kolaczyk, E. D., Dixon, D. D. 2000, \apj, 534, 490K
\bibitem[Kumar $\&$ Granot(2003)]{Kumar03} Kumar P., $\&$ Granot, J., 
\bibitem[Kumar et al.(2007)]{Kumar07} Kumar P. 2007, MNRAS, 376L, 57K
\bibitem[Liang et al.(2007)]{Liang07} Liang E., Zhang B., Zhang B., 2007, arXiv:0705.1373v2
\bibitem[Massaro, Conciatore, $\&$, Tramacere(2007)]{Massaro07} Massaro, F., Cutini, S., Conciatore, M. L., $\&$ Tramacere,  A. 2007, in prep.
\bibitem[M\'esz\'aros $\&$ Rees(1997)]{Meszaros97} M\'esz\'aros, P. $\&$ Rees M., 1997, ApJ, 476, 232
\bibitem[M\'esz\'aros(2002)]{Meszaros02} M\'esz\'aros, P. 2002 ARA$\&$A, 40, 137M
\bibitem[Mitsuda et al.(2007)]{Mitsuda07}, Mitsuda, K., et al. 2007, PASJ, 59, S1
\bibitem[Mundell et al.(2007)]{Mundell07} Mundell, C. 2007, ApJ, 660, 489M
\bibitem[Nousek et al.(2006)]{Nousek06} Nousek J. A. et al, 2006, ApJ, 642, 389 
\bibitem[Oates et al.(2007)]{Oates07} Oates, S. R., 2007, Accepted to MNRAS (arXiv0706.0669O)
\bibitem[Panaitescu $\&$ Kumar(2000)]{Panaitescu00} Panaitescu A., Kumar P., 2000, ApJ, 543, 66
\bibitem[Panaitescu et al.(2006)]{Panaitescu06} Panaitescu, A.; Meszaros, P.; Burrows, D.; Nousek, J.; Gehrels, N.; O'Brien, P.; Willingale, R. 2006, MNRAS, 369, 2059P
\bibitem[Panaitescu(2007a)]{Panaitescu07a} Panaitescu, A. 2007, accepted by MNRAS (2007arXiv0705.1015P)
\bibitem[Panaitescu(2007b)]{Panaitescu07b} Panaitescu, A. 2007, submitted to MNRAS (2007arXiv:0708.1509v2)
\bibitem[Perley et al.(2007)]{Perley07} Perley, D., et al. 2007, Submitted to ApJ (astro-ph/0703538)
\bibitem[Preece et al.(2000)]{Preece00} Preece, R. D., Briggs, M. S., Mallozzi, R. S., Pendleton, G. N., Paciesas, W. S. 2000, ApJS, 126, 19P
\bibitem[Rees $\&$ M\'esz\'aros(1998)]{Rees98} Rees, M. J., M\'esz\'aros, P., 1998, ApJ, 496, L1
\bibitem[Rhoads et al.(1997)]{Rhoads97} Rhoads, J. E. 1997, ApJ, 487, L1 
\bibitem[Sari, Piran $\&$ Narayan(1998)]{Sari98} Sari R., Piran T., Narayan R., 1998, ApJ, 497, L17
\bibitem[Sari et al.(1999)]{Sari99} Sari, R., Piran, T., $\&$ Halpern, J. P. 1999, ApJ, 519, L17
\bibitem[Sato et al.(2007)]{Sato07} Sato, G., 2007 ApJ, 657, 359S
\bibitem[Scargle(1998)]{Scargle98} Scargle, J. D. 1998, ApJ, 504, 405
\bibitem[Schady et al.(2006)]{Schady06} Schady, P., et al. 2006, Accepted by MNRAS (astro.ph.11081S)
\bibitem[Stanek et al.(1999)]{Stanek99} Stanek, K. Z., Garnavich, P. M., Kaluzny, J., Pych, W., $\&$ Thompson, I. 1999, ApJ, 522, L39 
\bibitem[Waxman, Kulkarni, $\&$ Frail(1998)]{Waxman98} Waxman, E., Kulkarni, S. R., $\&$ Frail, D. A. 1998, ApJ, 497, 288




\end{thebibliography}
\end{document}